\documentclass[3p]{elsarticle}

\usepackage[utf8]{inputenc}
\usepackage[T1]{fontenc}
\usepackage{uniinput}

\usepackage{hyperref}
\usepackage{amsmath,amssymb,amsfonts,bbm,dsfont,slashed,xfrac,nicefrac}
\usepackage{axodraw2}

\newcommand{\deq}{\mathrel{\mathop:}=}

\newcommand{\insertA}[1]{
\hspace{-4pt}
\raisebox{-.45\height}{
  \SetScale{#1}
  \begin{axopicture}(222,99)(99,-52)
    \SetWidth{1.5}
    \SetColor{Black}
    \Photon(150,1)(150,-51){3}{6}
    \Photon(200,1)(200,-51){3}{6}
    \Photon(125,-3)(125,21){3}{3}
    \Photon(270,1)(270,-51){3}{6}
    \SetWidth{2.0}
    \Line[arrow,arrowpos=0.125,arrowlength=7.5,arrowwidth=3,arrowinset=0,flip](100,-1)(200,-1)
    \Line(255,-1)(270,-1)
    \Line(200,-1)(215,-1)
    \Line[arrow,arrowpos=0.75,arrowlength=7.5,arrowwidth=3,arrowinset=0,flip](100,-1)(200,-1)
    \Line[arrow,arrowpos=0.37,arrowlength=7.5,arrowwidth=3,arrowinset=0,flip](100,-1)(200,-1)
    \Line[dash,dashsize=2](125,24)(125,49)
    \SetWidth{1.0}  
\Line[dash,dashsize=2,arrow,arrowpos=0.5,arrowlength=10,arrowwidth=4,arrowinset=0](125,27)(125,24)
    \Vertex(150,-1){3.606}
    \Vertex(200,-1){3.606}
    \Vertex(270,-1){3.606}
    \Vertex(125,-1){3.606}
    \Vertex(235,-1){2.236}
    \Vertex(245,-1){2.236}
    \SetWidth{2.0}
    \Line[arrow,arrowpos=0.5,arrowlength=7.5,arrowwidth=3,arrowinset=0,flip](270,-1)(320,-1)
    \SetWidth{1.0}
    \Vertex(225,-1){2.236}
  \end{axopicture}
 }
}
\newcommand{\insertB}[1]{
\hspace{-4pt}
\raisebox{-.45\height}{
  \SetScale{#1}
  \begin{axopicture}(222,99)(99,-52)
    \SetWidth{1.5}
    \SetColor{Black}
    \Photon(150,1)(150,-51){3}{6}
    \Photon(200,1)(200,-51){3}{6}
    \Photon(175,-3)(175,21){3}{3}
    \Photon(270,1)(270,-51){3}{6}
    \SetWidth{2.0}
    \Line[arrow,arrowpos=0.25,arrowlength=7.5,arrowwidth=3,arrowinset=0,flip](100,-1)(200,-1)
    \Line(255,-1)(270,-1)
    \Line[arrow,arrowpos=0.875,arrowlength=7.5,arrowwidth=3,arrowinset=0,flip](100,-1)(200,-1)
    \Line[arrow,arrowpos=0.5,arrowlength=7.5,arrowwidth=3,arrowinset=0,flip](270,-1)(320,-1)
    \Line[arrow,arrowpos=0.613,arrowlength=7.5,arrowwidth=3,arrowinset=0,flip](100,-1)(200,-1)
    \Line[dash,dashsize=2](175,24)(175,49)
    \SetWidth{1.0}
\Line[dash,dashsize=2,arrow,arrowpos=0.5,arrowlength=10,arrowwidth=4,arrowinset=0](175,27)(175,24)
    \Vertex(150,-1){3.606}
    \Vertex(200,-1){3.606}
    \Vertex(270,-1){3.606}
    \Vertex(175,-1){3.606}
    \SetWidth{2.0}
    \Line(200,-1)(215,-1)
    \SetWidth{1.0}
    \Vertex(225,-1){2.236}
    \Vertex(235,-1){2.236}
    \Vertex(245,-1){2.236}
  \end{axopicture}
 }
}
\newcommand{\insertC}[1]{
\hspace{-4pt}
\raisebox{-.45\height}{
  \SetScale{#1}
  \begin{axopicture}(222,99)(99,-52)
    \SetWidth{1.5}
    \SetColor{Black}
    \Photon(150,1)(150,-51){3}{6}
    \Photon(200,1)(200,-51){3}{6}
    \Photon(295,-3)(295,21){3}{3}
    \Photon(270,1)(270,-51){3}{6}
    \SetWidth{2.0}
    \Line[arrow,arrowpos=0.25,arrowlength=7.5,arrowwidth=3,arrowinset=0,flip](100,-1)(200,-1)
    \Line[arrow,arrowpos=0.75,arrowlength=7.5,arrowwidth=3,arrowinset=0,flip](100,-1)(200,-1)
    \Line[arrow,arrowpos=0.25,arrowlength=7.5,arrowwidth=3,arrowinset=0,flip](270,-1)(320,-1)
    \Line(255,-1)(270,-1)
    \Line(200,-1)(215,-1)
    \Line[dash,dashsize=2](295,24)(295,49)
    \SetWidth{1.0}
\Line[dash,dashsize=2,arrow,arrowpos=0.5,arrowlength=10,arrowwidth=4,arrowinset=0](295,27)(295,24)
    \Vertex(150,-1){3.606}
    \Vertex(200,-1){3.606}
    \Vertex(270,-1){3.606}
    \Vertex(295,-1){3.606}
    \SetWidth{2.0}
    \Line[arrow,arrowpos=0.75,arrowlength=7.5,arrowwidth=3,arrowinset=0,flip](270,-1)(320,-1)
    \SetWidth{1.0}
    \Vertex(225,-1){2.236}
    \Vertex(235,-1){2.236}
    \Vertex(245,-1){2.236}
  \end{axopicture}
 }
}
\newcommand{\slashL}[1]{
\hspace{-4pt}
\raisebox{-.45\height}{
  \SetScale{#1}
  \begin{axopicture}(222,99)(99,-52)
    \SetWidth{1.5}
    \SetColor{Black}
    \Photon(150,1)(150,-51){3}{6}
    \Photon(200,1)(200,-51){3}{6}
    \Photon(270,1)(270,-51){3}{6}
    \SetWidth{2.0}
    \Line[arrow,arrowpos=0.125,arrowlength=7.5,arrowwidth=3,arrowinset=0,flip](100,-1)(200,-1)
    \Line[arrow,arrowpos=0.375,arrowlength=7.5,arrowwidth=3,arrowinset=0,flip](100,-1)(200,-1)
    \Line[arrow,arrowpos=0.75,arrowlength=7.5,arrowwidth=3,arrowinset=0,flip](100,-1)(200,-1)
    \Line(200,-1)(215,-1)
    \Line[arrow,arrowpos=0.5,arrowlength=7.5,arrowwidth=3,arrowinset=0,flip](270,-1)(320,-1)
    \Line[dash,dashsize=2](125,-1)(125,49)
    \SetWidth{1.0}
    \Vertex(150,-1){3.606}
    \Vertex(200,-1){3.606}
    \Vertex(270,-1){3.606}
    \SetWidth{2.5}
    \Line(114,14)(114,-16)
    \SetWidth{1.0}
    \Vertex(225,-1){2.236}
    \Vertex(235,-1){2}
    \Vertex(245,-1){2.236}
    \SetWidth{2.0}
    \Line(255,-1)(270,-1)
  \end{axopicture}
 }
}
\newcommand{\slashR}[1]{
\hspace{-4pt}
\raisebox{-.45\height}{
  \SetScale{#1}
  \begin{axopicture}(222,99)(99,-52)
    \SetWidth{1.5}
    \SetColor{Black}
    \Photon(150,1)(150,-51){3}{6}
    \Photon(200,1)(200,-51){3}{6}
    \Photon(270,1)(270,-51){3}{6}
    \SetWidth{2.0}
    \Line[arrow,arrowpos=0.25,arrowlength=7.5,arrowwidth=3,arrowinset=0,flip](100,-1)(200,-1)
    \Line[arrow,arrowpos=0.75,arrowlength=7.5,arrowwidth=3,arrowinset=0,flip](270,-1)(320,-1)
    \Line(255,-1)(270,-1)
    \Line[arrow,arrowpos=0.75,arrowlength=7.5,arrowwidth=3,arrowinset=0,flip](100,-1)(200,-1)
    \Line(200,-1)(215,-1)
    \Line[dash,dashsize=2](295,-1)(295,49)
    \SetWidth{1.0}
    \Vertex(150,-1){3.606}
    \Vertex(200,-1){3.606}
    \Vertex(270,-1){3.606}
    \SetWidth{2.5}
    \Line(309,14)(309,-16)
    \SetWidth{1.0}
    \Vertex(225,-1){2.236}
    \Vertex(235,-1){2.236}
    \Vertex(245,-1){2.236}
    \SetWidth{2.0}
    \Line[arrow,arrowpos=0.25,arrowlength=7.5,arrowwidth=3,arrowinset=0,flip](270,-1)(320,-1)
  \end{axopicture}
 }
}
\newcommand{\exampleA}[1]{
\hspace{-4pt}
\raisebox{-.45\height}{
  \SetScale{#1}
  \begin{axopicture}(202,98)(74,-51)
    \SetWidth{1.5}
    \SetColor{Black}
    \Photon(100,0)(100,21.4){3}{3}
    \SetWidth{2.0}
    \Line[arrow,arrowpos=0.17,arrowlength=7.5,arrowwidth=3,arrowinset=0,flip](100,-2)(250,-2)
    \Line[arrow,arrowpos=0.5,arrowlength=7.5,arrowwidth=3,arrowinset=0,flip](100,-2)(250,-2)
    \Line[arrow,arrowpos=0.83,arrowlength=7.5,arrowwidth=3,arrowinset=0,flip](100,-2)(250,-2)
    \Line[arrow,arrowpos=0.4,arrowlength=7.5,arrowwidth=3,arrowinset=0,flip](75,-2)(100,-2)
    \Line[arrow,arrowpos=0.6,arrowlength=7.5,arrowwidth=3,arrowinset=0,flip](250,-2)(275,-2)
    \SetWidth{1.0}
\Line[dash,dashsize=2,arrow,arrowpos=0.5,arrowlength=10,arrowwidth=4,arrowinset=0](100,27)(100,24)
    \Vertex(150,-2){3.606}
    \Vertex(200,-2){3.606}
    \Vertex(250,-2){3.606}
    \Vertex(100,-2){3.606}
    \SetWidth{1.5}
    \Photon(150,0)(150,21.4){3}{3}
    \SetWidth{1.0}
\Line[dash,dashsize=2,arrow,arrowpos=0.5,arrowlength=10,arrowwidth=4,arrowinset=0](150,27)(150,24)
    \SetWidth{1.5}
    \PhotonArc(225,-11.054)(26,158.092,381.908){3}{15.5}
    \SetWidth{2.0}
    \Arc[dash,dashsize=2,clock](125,27.926)(25.074,-175.595,-364.405)
    \SetWidth{1.0}
    \Vertex(125,53){3.162}
  \end{axopicture}
 }
}
\newcommand{\exampleB}[1]{
\hspace{-4pt}
\raisebox{-.45\height}{
  \SetScale{#1}
  \begin{axopicture}(202,109)(74,-45)
    \SetWidth{2.0}
    \SetColor{Black}
    \Arc[dash,dashsize=2,clock](150,15.057)(53,156,22.843)
    \SetWidth{1.5}
    \Photon(100,11)(100,32.4){3}{3}
    \SetWidth{2.0}
    \Line[arrow,arrowpos=0.17,arrowlength=7.5,arrowwidth=3,arrowinset=0,flip](100,9)(250,9)
    \Line[arrow,arrowpos=0.5,arrowlength=7.5,arrowwidth=3,arrowinset=0,flip](100,9)(250,9)
    \Line[arrow,arrowpos=0.83,arrowlength=7.5,arrowwidth=3,arrowinset=0,flip](100,9)(250,9)
    \Line[arrow,arrowpos=0.4,arrowlength=7.5,arrowwidth=3,arrowinset=0,flip](75,9)(100,9)
    \Line[arrow,arrowpos=0.6,arrowlength=7.5,arrowwidth=3,arrowinset=0,flip](250,9)(275,9)
    \SetWidth{1.0}
\Line[dash,dashsize=2,arrow,arrowpos=0.5,arrowlength=10,arrowwidth=4,arrowinset=0](102,38)(101,35)
    \Vertex(150,9){3.606}
    \Vertex(200,9){3.606}
    \Vertex(250,9){3.606}
    \Vertex(100,9){3.606}
    \SetWidth{1.5}
    \PhotonArc(200,19.762)(50.762,-170.061,-9.939){3}{18.5}
    \Photon(200,11)(200,32.4){3}{3}
    \SetWidth{1.0}
\Line[dash,dashsize=2,arrow,arrowpos=0.5,arrowlength=10,arrowwidth=4,arrowinset=0](198,38)(199,35)
    \Vertex(150.5,68){3.162}
  \end{axopicture}
 }
}
\newcommand{\exampleC}[1]{
\hspace{-4pt}
\raisebox{-.45\height}{
  \SetScale{#1}
  \begin{axopicture}(202,109)(74,-45)
    \SetWidth{2.0}
    \SetColor{Black}
    \Arc[dash,dashsize=2,clock](174.5,-21.083)(94.085,141.583,38.417)
    \SetWidth{1.5}
    \Photon(100,11)(100,32.4){3}{3}
    \SetWidth{2.0}
    \Line[arrow,arrowpos=0.17,arrowlength=7.5,arrowwidth=3,arrowinset=0,flip](100,9)(250,9)
    \Line[arrow,arrowpos=0.5,arrowlength=7.5,arrowwidth=3,arrowinset=0,flip](100,9)(250,9)
    \Line[arrow,arrowpos=0.83,arrowlength=7.5,arrowwidth=3,arrowinset=0,flip](100,9)(250,9)
    \Line[arrow,arrowpos=0.4,arrowlength=7.5,arrowwidth=3,arrowinset=0,flip](75,9)(100,9)
    \Line[arrow,arrowpos=0.6,arrowlength=7.5,arrowwidth=3,arrowinset=0,flip](250,9)(275,9)
    \SetWidth{1.0}
\Line[dash,dashsize=2,arrow,arrowpos=0.5,arrowlength=10,arrowwidth=4,arrowinset=0](102,38)(101,35)
    \Vertex(150,9){3.606}
    \Vertex(200,9){3.606}
    \Vertex(250,9){3.606}
    \Vertex(100,9){3.606}
    \SetWidth{1.5}
    \PhotonArc(175,-0.054)(26,158.092,381.908){3}{13.5}
    \Photon(250,11)(249,32.4){3}{3}
    \SetWidth{1.0}
\Line[dash,dashsize=2,arrow,arrowpos=0.5,arrowlength=10,arrowwidth=4,arrowinset=0](247,38)(248,35)
    \Vertex(174.5,73.5){3.162}
  \end{axopicture}
 }
}
\newcommand{\exampleD}[1]{
\hspace{-4pt}
\raisebox{-.45\height}{
  \SetScale{#1}
  \begin{axopicture}(202,116)(74,-41)
    \SetWidth{1.5}
    \SetColor{Black}
    \Photon(200,18)(200,39.4){3}{3}
    \SetWidth{2.0}
    \Line[arrow,arrowpos=0.17,arrowlength=7.5,arrowwidth=3,arrowinset=0,flip](100,16)(250,16)
    \Line[arrow,arrowpos=0.5,arrowlength=7.5,arrowwidth=3,arrowinset=0,flip](100,16)(250,16)
    \Line[arrow,arrowpos=0.83,arrowlength=7.5,arrowwidth=3,arrowinset=0,flip](100,16)(250,16)
    \Line[arrow,arrowpos=0.4,arrowlength=7.5,arrowwidth=3,arrowinset=0,flip](75,16)(100,16)
    \Line[arrow,arrowpos=0.6,arrowlength=7.5,arrowwidth=3,arrowinset=0,flip](250,16)(275,16)
    \SetWidth{1.0}
\Line[dash,dashsize=2,arrow,arrowpos=0.5,arrowlength=10,arrowwidth=4,arrowinset=0](200,45)(200,42)
    \Vertex(150,16){3.606}
    \Vertex(200,16){3.606}
    \Vertex(250,16){3.606}
    \Vertex(100,16){3.606}
    \SetWidth{1.5}
    \Photon(150,18)(150,39.4){3}{3}
    \SetWidth{1.0}
\Line[dash,dashsize=2,arrow,arrowpos=0.5,arrowlength=10,arrowwidth=4,arrowinset=0](150,45)(150,42)
    \SetWidth{1.5}
    \PhotonArc(175,41.73)(78.664,-162.443,-17.557){3}{28.5}
    \SetWidth{2.0}
    \Arc[dash,dashsize=2](175,45.926)(25.074,-4.405,184.405)
    \SetWidth{1.0}
    \Vertex(175,71){3.162}
  \end{axopicture}
 }
}
\newcommand{\exampleE}[1]{
\hspace{-4pt}
\raisebox{-.45\height}{
  \SetScale{#1}
  \begin{axopicture}(202,109)(74,-45)
    \SetWidth{2.0}
    \SetColor{Black}
    \Arc[dash,dashsize=2,clock](199.283,15.057)(53,156,22.843)
    \SetWidth{1.5}
    \Photon(150,11)(150,32.4){3}{3}
    \SetWidth{2.0}
    \Line[arrow,arrowpos=0.17,arrowlength=7.5,arrowwidth=3,arrowinset=0,flip](100,9)(250,9)
    \Line[arrow,arrowpos=0.5,arrowlength=7.5,arrowwidth=3,arrowinset=0,flip](100,9)(250,9)
    \Line[arrow,arrowpos=0.83,arrowlength=7.5,arrowwidth=3,arrowinset=0,flip](100,9)(250,9)
    \Line[arrow,arrowpos=0.4,arrowlength=7.5,arrowwidth=3,arrowinset=0,flip](75,9)(100,9)
    \Line[arrow,arrowpos=0.6,arrowlength=7.5,arrowwidth=3,arrowinset=0,flip](250,9)(275,9)
    \SetWidth{1.0}
\Line[dash,dashsize=2,arrow,arrowpos=0.5,arrowlength=10,arrowwidth=4,arrowinset=0](152,38)(151,35)
    \Vertex(150,9){3.606}
    \Vertex(200,9){3.606}
    \Vertex(250,9){3.606}
    \Vertex(100,9){3.606}
    \SetWidth{1.5}
    \PhotonArc(150,19.762)(50.762,-170.061,-9.939){3}{18.5}
    \Photon(250,11)(250,32.4){3}{3}
    \SetWidth{1.0}
\Line[dash,dashsize=2,arrow,arrowpos=0.5,arrowlength=10,arrowwidth=4,arrowinset=0](248,38)(249,35)
    \Vertex(200,68.5){3.162}
  \end{axopicture}
 }
}
\newcommand{\exampleF}[1]{
\hspace{-4pt}
\raisebox{-.45\height}{
  \SetScale{#1}
  \begin{axopicture}(202,98)(74,-51)
    \SetWidth{1.5}
    \SetColor{Black}
    \Photon(200,0)(200,21.4){3}{3}
    \SetWidth{2.0}
    \Line[arrow,arrowpos=0.17,arrowlength=7.5,arrowwidth=3,arrowinset=0,flip](100,-2)(250,-2)
    \Line[arrow,arrowpos=0.5,arrowlength=7.5,arrowwidth=3,arrowinset=0,flip](100,-2)(250,-2)
    \Line[arrow,arrowpos=0.83,arrowlength=7.5,arrowwidth=3,arrowinset=0,flip](100,-2)(250,-2)
    \Line[arrow,arrowpos=0.4,arrowlength=7.5,arrowwidth=3,arrowinset=0,flip](75,-2)(100,-2)
    \Line[arrow,arrowpos=0.6,arrowlength=7.5,arrowwidth=3,arrowinset=0,flip](250,-2)(275,-2)
    \SetWidth{1.0}
\Line[dash,dashsize=2,arrow,arrowpos=0.5,arrowlength=10,arrowwidth=4,arrowinset=0](200,27)(200,24)
    \Vertex(150,-2){3.606}
    \Vertex(200,-2){3.606}
    \Vertex(250,-2){3.606}
    \Vertex(100,-2){3.606}
    \SetWidth{1.5}
    \PhotonArc(125,-11.054)(26,158.092,381.908){3}{13.5}
    \Photon(250,0)(250,21.4){3}{3}
    \SetWidth{1.0}
\Line[dash,dashsize=2,arrow,arrowpos=0.5,arrowlength=10,arrowwidth=4,arrowinset=0](250,27)(250,24)
    \SetWidth{2.0}
    \Arc[dash,dashsize=2,clock](225,27.926)(25.074,-175.595,-364.405)
    \SetWidth{1.0}
    \Vertex(225,53){3.162}
  \end{axopicture}
 }
}
\newcommand{\tadA}[1]{
\hspace{-2pt}
\raisebox{-.45\height}{
  \SetScale{#1}
  \begin{axopicture}(202,78)(74,-61)
    \SetWidth{2.0}
    \SetColor{Black}
    \Arc[dash,dashsize=2,clock](125,-13.429)(26.429,-161.075,-378.925)
    \Line[arrow,arrowpos=0.17,arrowlength=7.5,arrowwidth=3,arrowinset=0,flip](100,-22)(250,-22)
    \Line[arrow,arrowpos=0.5,arrowlength=7.5,arrowwidth=3,arrowinset=0,flip](100,-22)(250,-22)
    \Line[arrow,arrowpos=0.83,arrowlength=7.5,arrowwidth=3,arrowinset=0,flip](100,-22)(250,-22)
    \Line[arrow,arrowpos=0.4,arrowlength=7.5,arrowwidth=3,arrowinset=0,flip](75,-22)(100,-22)
    \Line[arrow,arrowpos=0.6,arrowlength=7.5,arrowwidth=3,arrowinset=0,flip](250,-22)(275,-22)
    \SetWidth{1.0}
    \Vertex(200,-22){3.606}
    \Vertex(250,-22){3.606}
    \SetWidth{1.5}
    \PhotonArc(225,-31.054)(26,158.092,381.908){3}{15.5}
    \SetWidth{1.0}
    \Vertex(125,13){3.162}
    \SetWidth{2.5}
    \Line(127,-7)(127,-37)
    \Line(87,-7)(87,-37)
  \end{axopicture}
 }
}
\newcommand{\exampleG}[1]{
\hspace{-4pt}
\raisebox{-.45\height}{
  \SetScale{#1}
  \begin{axopicture}(202,103)(74,-48)
    \SetWidth{2.0}
    \SetColor{Black}
    \Line[arrow,arrowpos=0.17,arrowlength=7.5,arrowwidth=3,arrowinset=0,flip](100,3)(250,3)
    \Line[arrow,arrowpos=0.5,arrowlength=7.5,arrowwidth=3,arrowinset=0,flip](100,3)(250,3)
    \Line[arrow,arrowpos=0.83,arrowlength=7.5,arrowwidth=3,arrowinset=0,flip](100,3)(250,3)
    \Line[arrow,arrowpos=0.4,arrowlength=7.5,arrowwidth=3,arrowinset=0,flip](75,3)(100,3)
    \Line[arrow,arrowpos=0.6,arrowlength=7.5,arrowwidth=3,arrowinset=0,flip](250,3)(275,3)
    \SetWidth{1.0}
    \Vertex(150,3){3.606}
    \Vertex(200,3){3.606}
    \SetWidth{1.5}
    \PhotonArc(175,-6.054)(26,158.092,381.908){3}{13.5}
    \SetWidth{2.0}
    \Arc[dash,dashsize=2,clock](175,-13.875)(76.875,167.32,12.68)
    \SetWidth{1.0}
    \Vertex(175,63){2.828}
    \SetWidth{2.5}
    \Line(87,18)(87,-12)
    \Line(266,18)(266,-12)
  \end{axopicture}
 }
}
\newcommand{\tadB}[1]{
\hspace{-2pt}
\raisebox{-.45\height}{
  \SetScale{#1}
  \begin{axopicture}(202,78)(74,-61)
    \SetWidth{2.0}
    \SetColor{Black}
    \Arc[dash,dashsize=2,clock](225,-13.429)(26.429,-161.075,-378.925)
    \Line[arrow,arrowpos=0.17,arrowlength=7.5,arrowwidth=3,arrowinset=0,flip](100,-22)(250,-22)
    \Line[arrow,arrowpos=0.5,arrowlength=7.5,arrowwidth=3,arrowinset=0,flip](100,-22)(250,-22)
    \Line[arrow,arrowpos=0.83,arrowlength=7.5,arrowwidth=3,arrowinset=0,flip](100,-22)(250,-22)
    \Line[arrow,arrowpos=0.4,arrowlength=7.5,arrowwidth=3,arrowinset=0,flip](75,-22)(100,-22)
    \Line[arrow,arrowpos=0.6,arrowlength=7.5,arrowwidth=3,arrowinset=0,flip](250,-22)(275,-22)
    \SetWidth{1.0}
    \Vertex(150,-22){3.606}
    \Vertex(100,-22){3.606}
    \SetWidth{1.5}
    \PhotonArc(125,-31.054)(26,158.092,381.908){3}{13.5}
    \SetWidth{1.0}
    \Vertex(225,13){3.162}
    \SetWidth{2.5}
    \Line(225,-7)(225,-37)
    \Line(266,-7)(266,-37)
  \end{axopicture}
 }
}
\newcommand{\DSE}[1]{
\raisebox{-.35\height}{
  \SetScale{#1}
  \begin{axopicture}(202,85)(74,-36)
    \SetWidth{2.0}
    \SetColor{Black}
    \EBox(150,-35)(200,5)
    \Line[arrow,arrowpos=0.5,arrowlength=7.5,arrowwidth=3,arrowinset=0,flip](100,-15)(150,-15)
    \Line[arrow,arrowpos=0.5,arrowlength=7.5,arrowwidth=3,arrowinset=0,flip](200,-15)(250,-15)
    \Line[arrow,arrowpos=0.4,arrowlength=7.5,arrowwidth=3,arrowinset=0,flip](75,-15)(100,-15)
    \Line[arrow,arrowpos=0.6,arrowlength=7.5,arrowwidth=3,arrowinset=0,flip](250,-15)(275,-15)
    \Arc[dash,dashsize=2,clock](175,-31.875)(76.875,167.32,12.68)
    \SetWidth{1.0}
    \Vertex(175,45){2.828}
    \SetWidth{2.5}
    \Line(87,0)(87,-30)
    \Line(266,0)(266,-30)
    \Text(159,-26)[lb]{\Black{$c_{n;l}$}}
  \end{axopicture}
 }
}

\newcommand{\treeA}[1]{
\raisebox{-.45\height}{
  \SetScale{#1}
  \begin{axopicture}(102,53)(74,-52)
    \SetWidth{1.5}
    \SetColor{Black}
    \Photon(125,-49)(125,-25){3}{3}
    \SetWidth{2.0}
    \Line[arrow,arrowpos=0.5,arrowlength=7.5,arrowwidth=3,arrowinset=0,flip](125,-47)(175,-47)
    \Line[arrow,arrowpos=0.5,arrowlength=7.5,arrowwidth=3,arrowinset=0,flip](75,-47)(125,-47)
    \Line[dash,dashsize=2](125,-22)(125,3)
    \SetWidth{1.0}
\Line[dash,dashsize=2,arrow,arrowpos=0.5,arrowlength=10,arrowwidth=4,arrowinset=0](125,-19)(125,-23)
    \Vertex(125,-47){3.606}
  \end{axopicture}
 }
}
\newcommand{\treeB}[1]{
\raisebox{-.45\height}{
  \SetScale{#1}
  \begin{axopicture}(102,53)(74,-52)
    \SetWidth{2.0}
    \SetColor{Black}
    \Line[arrow,arrowpos=0.5,arrowlength=7.5,arrowwidth=3,arrowinset=0,flip](125,-47)(175,-47)
    \Line[arrow,arrowpos=0.5,arrowlength=7.5,arrowwidth=3,arrowinset=0,flip](75,-47)(125,-47)
    \Line[dash,dashsize=2](125,-47)(125,3)
    \SetWidth{2.5}
    \Line(100,-32)(100,-62)
  \end{axopicture}
 }
}
\newcommand{\treeC}[1]{
\raisebox{-.45\height}{
  \SetScale{#1}
  \begin{axopicture}(102,53)(74,-52)
    \SetWidth{2.0}
    \SetColor{Black}
    \Line[arrow,arrowpos=0.5,arrowlength=7.5,arrowwidth=3,arrowinset=0,flip](125,-47)(175,-47)
    \Line[arrow,arrowpos=0.5,arrowlength=7.5,arrowwidth=3,arrowinset=0,flip](75,-47)(125,-47)
    \Line[dash,dashsize=2](125,-47)(125,3)
    \SetWidth{2.5}
    \Line(150,-32)(150,-62)
  \end{axopicture}
 }
}

\begin{document}

\begin{frontmatter}
 
\title{Diagrammatic Cancellations and the Gauge Dependence of QED}

\author[a,b]{Henry Kißler\corref{cor1}}
\ead{kissler@physik.hu-berlin.de}
\address[a]{Department of Mathematical Sciences, University of Liverpool, L69 7ZL, Liverpool, 
United Kingdom}
\address[b]{Department of Mathematics, Humboldt-Universität zu Berlin, Rudower Chaussee 25, 
D-12489 Berlin, Germany}

\author[b]{Dirk Kreimer}
\ead{kreimer@math.hu-berlin.de}

\cortext[cor1]{Corresponding author}

\begin{abstract}
This letter examines diagrammatic cancellations for Quantum Electrodynamics (QED) in the general 
linear gauge. These cancellations combine Feynman graphs of various topologies and provide a method 
to reconstruct the gauge dependence of the electron propagator from the result of a particular 
gauge by means of a linear Dyson-Schwinger equation. We use this method in combination with 
dimensional regularization to demonstrate how the 3-loop $ε$-expansion in the Feynman gauge 
determines the $ε$-expansions for all gauge parameter dependent terms to 4 loops.
\end{abstract}

 
\end{frontmatter}

\section{Introduction}

Our ability to perform high-order calculations in perturbative Quantum Field Theory is limited 
due to the enormous number of Feynman graphs which need to be evaluated. A graph-by-graph 
evaluation fails to exploit global symmetry properties, such as the Ward or Slavnov-Taylor 
identities, which are only satisfied by the sum of a certain subclass of Feynman graphs. 
However, for abelian as well as non-abelian gauge theories, it is well understood that these global 
identities originate from the pure structure of propagator and vertex Feynman rules and their 
resulting cancellations between certain tree-level Feynman graphs  
\cite{'tHooft:1971fh,'tHooft:1972ue,Lautrup:1977ty,Cvitanovic:1977qu,Kreimer:2012jw,Sars:2015phd}. 
Although these combinatorial arguments are considered to be rather inconvenient and lengthy in 
comparison to a BRST derivation, there are at least three potential benefits for exploiting these 
cancellations in perturbative Quantum Field Theory.
\begin{itemize}
 \item The perturbative expansion given in terms of Feynman graphs might be rearranged in terms of 
  meta graphs or subsectors with a maximum number of cancellations implemented. For example, an 
  early attempt to construct gauge invariant subsectors in QCD was given by Cvitanović et al.\ 
  \cite{Cvitanovic:1980bu}.
 \item An explicit implementation of a cancellation identity into computational procedures reduces 
  the number of terms which need to be evaluated, for example Herzog et al.\ pointed out that 
  contracting all legs of a three-gluon vertex with the corresponding momenta yields a vanishing 
  contribution \cite{Herzog:2016qas,Ueda:2016sxw}. Similar more general identities on gauge 
  theory amplitudes lead to the graph and cycle homologies observed in \cite{Kreimer:2012jw}.
  \item Cancellation identities provide all-order restrictions for the structure of Green's 
  function, for instance certain tree-level identities guarantee the transversality of the photon 
  or gluon self-energy. This can be exploited to simplify calculations or provide a non-trivial 
  check.
\end{itemize}

This letter provides an insight into how the QED tree-level identity implies cancellation between 
Feynman graphs of different topologies and determines the gauge dependence.

\section{The QED cancellation identity}

This section discusses the QED cancellation identity and its implications. Special emphasis is given 
to the electron propagator.

Consider the Lagrangian of Quantum Electrodynamics in the linear covariant gauge with fermions of 
mass $m$ 
\begin{align}
  L_\text{QED} =  - \frac{1}{4} F_{μν}F^{μν} -\frac{(\partial_μ A^μ)^2}{2 (1-ξ)}  + 
\bar{ψ}(i\slashed{D} -m) ψ.
  \label{eq:qedLagrangian}
\end{align}
This parametrization of the gauge parameter $ξ$ is especially convenient for perturbative 
calculations as it minimizes the number of terms 
in the photon propagator
\begin{align}
  P^{μν} (k) = -i \left(\frac{g^{μν}}{k^2} - ξ \frac{k^μ k^ν}{(k^2)^2}\right).
  \label{eq:photonProp}
\end{align}
A generic Feynman graph evaluates to a polynomial in $ξ$. The constant term of this polynomial is 
derived by setting all photon propagators in the Feynman gauge ($ξ=0$). Terms of higher powers in 
$ξ$ can be constructed by replacements of these Feynman-gauged propagators by the gauge dependent 
term of the photon propagator. For instance, the term linear in $ξ$ corresponds to the summation 
over all possibilities to replace one of the photon propagators which are in the Feynman gauge by a 
gauge dependent Lorentz tensor
\begin{align}
 -k^μ \frac{1}{k^2} (-i ξ) \frac{1}{k^2}\, k^ν.
 \label{eq:gaugeTensor}
\end{align}
In a Feynman graph, this tensor connects two electron-photon vertices and both of these vertices 
are contracted with their ingoing photon momentum. The result of such a contraction is the content 
of the famous tree-level identity \cite{Bjorken:1965zz}
\begin{align}
  \begin{tabular}{ccccc}
$\!{\begin{aligned}\frac{1}{\slashed{p}+\slashed{k}-m}γ_νk^ν\frac{1}{\slashed{p}-m}\end{aligned}}$ &
$=$ &    $\!{\begin{aligned}\frac{1}{\slashed{p}-m}\end{aligned}}$ & 
$-$ &    $\!{\begin{aligned}\frac{1}{\slashed{p}+\slashed{k}-m},\end{aligned}}$\\
$\treeA{.45}$ & $=$ & $\treeB{.45}$ & $-$ & $\treeC{.45}$.
  \end{tabular}
  \label{eq:treeLevel1}
\end{align}
The second line introduces some graphical notation: the contraction of the vertex with its ingoing 
photon momentum is represented by attaching a triangle on top of the wavy photon line, cancelled 
fermion propagators are visualized by slashed fermion lines, the remaining dashed line inserts
the photon momentum at the vertex, but without contributing the usual $γ^ν$ vertex factor.

The successive iteration of this tree-level identity implies the QED cancellation identity
\begin{align}
 \insertA{.35} + \insertB{.35} + \cdots + \insertC{.35} = \slashL{.35} - \slashR{.35}.
  \label{eq:treeLevel2}
\end{align}
It should be remarked that there is no restriction on the photon lines below the horizontal fermion 
line --- pairs of them might be connected with propagators in the Feynman gauge, the gauge 
dependent Lorentz tensor \eqref{eq:gaugeTensor}, or vacuum polarization graphs. This feature 
enables the tree-level identity to describe cancellations between Feynman graphs at a particular 
loop order. 

As a result, these tree-level identities guarantee the Ward-Takahasih identity.

For instance, the transversality of the vacuum polarization follows from equation 
\eqref{eq:treeLevel2} by closing the horizontal electron line to a fermion loop and using the 
momentum routing invariance such that the right-hand side of the equation vanishes. The second 
consequence of an iterative application of suchlike cancellations implies that the summation over 
all insertion places of the tensor \eqref{eq:gaugeTensor} into all vacuum polarization graphs 
yields a vanishing contribution, i.e.\ the vacuum polarization is independent of the gauge 
parameter.

Further, it is possible to give a diagrammatic derivation \cite{'tHooft:1971fh,Cvitanovic:1983eb} of 
the identity
\begin{align}
  k^μ Λ_μ (k,p) = Σ(p) - Σ(p+k)
\end{align}
which relates the amputated vertex function $Λ_μ$ to the self-energy of the electron $Σ$. 
Therefore, we might concentrate our discussion on the latter.

Indeed, all gauge dependent Feynman graphs of the connected electron propagator function $S$ can be 
constructed from this cancellation identity. In order to derive the term linear in the gauge 
parameter $ξ$ it is necessary to sum over all possible insertions of the gauge dependent tensor 
\eqref{eq:gaugeTensor}. This corresponds to attaching a second momentum-contracted photon leg on 
the horizontal fermion line in the cancellation identity \eqref{eq:treeLevel2}. After summation over 
all possible insertion places, the cancellation identity can be iteratively used to 
reduce all momentum-contracted photon lines to dashed lines, which are attached to the ends of the 
horizontal fermion line.

An application of this procedure produces the following cancellation between Feynman graphs at two 
loops and linear in $ξ$.
\begin{multline}
  \exampleA{.35} + \exampleB{.35} + \exampleC{.35} + \exampleD{.35} + \exampleE{.35} + 
  \exampleF{.35}\\
  = \frac{1}{2} \tadA{.35} - \exampleG{.35} + \frac{1}{2} \tadB{.35}
\end{multline}
Here, the scalar part of the gauge dependent tensor \eqref{eq:gaugeTensor} is represented by the 
dashed propagator lines which enclose a dot to denote the $ξ$ factor. As seen by the first graph, 
the summation over all insertion places requires us to consider \emph{connected} rather then 
one-particle irreducible Feynman graphs. The two tadpole graphs are weighted by a combinatorial 
factors of $\nicefrac{1}{2}$, which originates from the fact that the summation over all insertion 
places overcounts graphs which are symmetric under interchange of the momentum-contracted photon 
legs. Remarkably, this two-loop example demonstrates how Feynman graphs of the T1, T2, and 
T3 {\sc Mincer} topologies \cite{Larin:1991fz} combine and beside the vanishing tadpoles, only 
a T2 topology graph remains. Eventually, the two-loop gauge dependent term is constructed by 
inserting the one-loop propagator 
into a one-loop skeleton graph.

This observation generalizes to the full connected electron propagator by means of a linear 
Dyson-Schwinger equation. The electron propagator is expanded into coefficients $c_{n;l}$ of 
loop order $n$ and order $l$ in the gauge parameter
\begin{align}
  S(q,m) &= \frac{i}{\slashed{q}-m} + \sum_{1\leq n} \sum_{0\leq l \leq n} c_{n;l} ξ^l α^n.
  \label{eq:connected}
\end{align}
These coefficients depend on the momentum $q$ and the mass $m$, $c_{n+1;l+1}\left(q,m\right)$, and 
can be constructed by attaching $(l+1)$ pairs of momentum-contracted photon legs on the horizontal 
fermion line and connecting the remaining photon legs by the Feynman-gauged propagator or vacuum 
polarization graphs in all possible ways. The summation over all possible insertion places of the 
momentum-contracted photon legs allows for the iterated application of the cancellation identity 
\eqref{eq:treeLevel2} and implies the linear Dyson-Schwinger equation
\begin{align}
 c_{n+1; l+1}(q,m) = - \frac{1}{l+1} e^2 ξ \int \frac{d^4k}{(2π)^4}  \frac{-i}{\left[-k^2\right]^2} 
c_{n;l}\left(q+k,m\right) = - \frac{1}{l+1}\DSE{.45}.
 \label{eq:dse}
\end{align}
This reduction can be iterated and is valid at an arbitrary loop order. The factor 
$\nicefrac{1}{(l+1)}$ compensates the overcounting generated by the summation over all insertion 
places or, in other words, it adjusts for the fact that each of the $(l+1)$ gauge dependent 
propagators contribute one term to the cancellation identity. Interpreting this factor as the 
derivative of $c_{n+1; l+1}$ with respect to the gauge parameter, our result shows a remarkable 
similarity to the famous Landau-Khalatnikov formula 
\cite{Landau:1955zz,Fradkin:1955jr,Johnson:1959zz}, or rather a recently derived momentum space 
version \cite{Jia:2016udu,Jia:2016wyu}. For further discussions of the gauge dependence in Quantum 
Electrodynamics and Quantum Chromodynamics based on other methods but related to the 
Landau-Khalatnikov formula the reader is referred to \cite{Sonoda:2000kn,Nishijima:2007ry}.

\section{Gauge-dependent terms to 4 loops}

This section focuses on the massless limit of Quantum Electrodynamics. In this limit, the bare 
electron propagator is analysed by means of the Dyson-Schwinger equation \eqref{eq:dse} and 
compared to results obtained by perturbative computations using dimensional regularization.

The skeleton graph of the Dyson-Schwinger equation can be written in terms of the dimensional 
regularized one-loop master integral of two scalar propagators with weights $x$ and $y$
\begin{align}
  G(x,y) &= -i \int \frac{d^Dk}{(2π)^D}
  \frac{(-q^2)^{x+y-D/2}}{\left[-k^2\right]^x\left[-(k+q)^2\right]^y}
  = \frac{Γ(\nicefrac{D}{2}-x) Γ(\nicefrac{D}{2}-y)Γ(x+y-\nicefrac{D}{2})}{(4π)^{\nicefrac{D}{2}} 
  Γ(x) Γ(y) Γ(D-x-y)}
\end{align}
\cite{Chetyrkin:1980pr}, where we have amputated the dependence on the external momentum for 
notational convenience. Recall 
that the momentum dependence of the $n$-loop bare electron propagator $c_{n;l}(q)$ is given by
$\nicefrac{\slashed{q}}{(q^2)^{1+nε}}$. Insertion of this momentum dependence into the 
Dyson-Schwinger equation determines the contribution of the skeleton graph to the electron 
propagator. In $D = d-2ε$ dimensions, it contributes the factor
\begin{align}
 F(d,ε,n) = \frac{1}{2}e^2ξ \left(-q^2\right)^{-ε}
 \left[ G(\nicefrac{d}{2}-1,1+nε) - G(\nicefrac{d}{2},1+nε)- G(\nicefrac{d}{2},nε) \right]
 \exp\left[ε \left(γ_E+ \dfrac{ζ(2) ε}{2}\right)\right].
 \label{eq:skeletonF}
\end{align}
Here, $e$ denotes the bare coupling parameter, $q$ the external momentum of the electron line, 
$γ_E$ the Euler–Mascheroni constant and $ζ(z)$ is the Riemann zeta function. The exponential factor 
does not originate from the Dyson-Schwinger equation, but is implemented for straightforward 
comparison with {\sc Mincer} results. So far, we are only concerned with the four dimensional case 
$(d=4)$ and hence define $F(ε,n) \deq F(4,ε,n)$ and set $c_{0;0} \deq \nicefrac{i}{\slashed{q}}$. 
Now, 
an iterative use of the Dyson-Schwinger equation determines all coefficients 
\begin{align}
 c_{n;l} = \frac{1}{l!} c_{n-l;0} \prod_{1 \leq j \leq l} F(ε,n-j) \text{ for }l \geq 1,
 \label{eq:coeffRecursion}
\end{align}
in terms of the Feynman gauge result $c_{n;0}$ and the one-loop skeleton function $F$. For 
instance, the $ε$-expansions of all gauge dependent terms at 4 loops 
\begin{align}
\label{eq:fourLoop1}
  c_{4;1}(ε)& = c_{3;0}(ε) F(ε,3),\\
\label{eq:fourLoop2}
  c_{4;2}(ε)& = \frac{1}{2!} c_{2;0}(ε) F(ε,3) F(ε,2),\\
\label{eq:fourLoop3}
  c_{4;3}(ε)& = \frac{1}{3!} c_{1;0}(ε) F(ε,3) F(ε,2) F(ε,1),\\
\label{eq:fourLoop4}
  c_{4;4}(ε)& = \frac{1}{4!} c_{0;0} F(ε,3)F(ε,2)F(ε,1)F(ε,0)
\end{align}
are determined once the $ε$-expansion is known in Feynman gauge at 3 loops. For analytic results of 
the electron propagator to three loops the reader is referred to 
\cite{Johnson:1964da,Tarasov:1980au,Larin:1993tp,Broadhurst:1995dq}.

Recall that these coefficients refer to the connected electron propagator \eqref{eq:connected}. 
However, for the sake of computational convenience, it is necessary to relate these coefficients to 
the one-particle irreducible self-energy of the electron $Σ$. Define coefficients $p_{n;l}$ to 
decompose the self-energy
\begin{align}
 Σ(q) & = \slashed{q} \sum_{1\leq n} \sum_{0\leq l \leq n} p_{n;l} ξ^l α^n .
\end{align}
Then, the relation between the connected and the one-particle irreducible propagator
\begin{align}
 S(q) & = \frac{i}{\slashed{q} - Σ(q)}
\intertext{implies the conversion formulas for the coefficients}
  \tilde{c}_{n;l} & \deq -i \slashed{q} c_{n;l} = \sum_{k\geq1} 
\sum_{\substack{n_1+\cdots+n_k=n\\l_1+\cdots+l_k=l}} p_{n_1;l_1}   \cdots 
p_{n_k;l_k},\label{eq:pIntoC}\\
  p_{n;l} &= \sum_{k\geq1} (-1)^{k+1} \sum_{\substack{n_1+\cdots+n_k=n\\l_1+\cdots+l_k=l}} 
  \tilde{c}_{n_1;l_1} \cdots \tilde{c}_{n_k;l_k}.
  \label{cIntoP}
\end{align}

In this way, we are able to compare the formula for the coefficient with results from actual 
perturbative computations. We generate the Feynman graphs of the electron self-energy $Σ$ with {\sc 
Qgraf} \cite{Nogueira:1991ex} and perform the computations in {\sc FORM} \cite{Vermaseren:2000nd} 
and its parallel version {\sc TFORM} \cite{Tentyukov:2007mu} in combination with the {\sc Mincer} 
package \cite{Gorishnii:1989gt,Larin:1991fz}. From these results, we extract the coefficients 
$p_{n;m}$ as a series in $ε$ to three loops $n \leq 3$ with an arbitrary power in the gauge 
parameter $0 \leq l \leq n$. After the conversion into the connected coefficient \eqref{eq:pIntoC}, 
the {\sc Mincer} expansions exactly match the formula \eqref{eq:coeffRecursion} derived by means of 
the linear Dyson-Schwinger equation.

\section{Conclusions and outlook}

In this letter, we utilized tree-level identities to examine diagrammatic cancellations in QED. In 
general, these cancellations involve planar as well non-planar Feynman graphs. Our method applies 
to an arbitrary order in perturbation theory and allowed us to construct the gauge dependence of the 
electron propagator to 4 loops from its value in the Feynman gauge. Similar reconstructions are 
possible if the electron propagator is known in a different linear covariant gauge --- we 
explicitly checked this for the Landau gauge. 

Remarkably, the result of a specific gauge at particular loop order determines all gauge terms 
even at the next loop order and the recursion \eqref{eq:coeffRecursion} provides a closed-form 
expression of the coefficients proportional to $α^n ξ^n$ at an arbitrary loop order $n$.

These results follow from the linear Dyson-Schwinger equation \eqref{eq:dse}, which 
might be of interest for non-perturbative studies. However, the dipole scalar propagator 
introduces an infrared singularity in the kernel of the skeleton. In contrast to usual ultraviolet 
divergences, this infrared singularity is not regularized by insertion of an ansatz for the 
renormalized electron propagator into the Dyson-Schwinger equation. A suitable method to construct 
a renormalized Dyson-Schwinger equation might be the implementation of cancellation identities into 
the Hopf algebra of QED \cite{vanSuijlekom:2006ig,Kissler:2016gxn}.

Recently, gauge theories in higher dimensions attracted great interest, because of the 
interconnection of infrared and ultraviolet fixed points of various dimensions 
\cite{Kazakov:2002jd,Kazakov:2007bw}. At $d=4,6,8$ the photon propagator possesses the same 
Lorentz tensor structure as in \eqref{eq:photonProp} and its gauge dependent tensor has a scalar 
factor of ${(k^2)}^{\nicefrac{d}{2}}$ in the denominator \cite{Gracey:2015xmw}. Therefore, the 
expression for the skeleton \eqref{eq:skeletonF} can also be used to construct the gauge dependence 
in $d=6$ dimensions. However, the eight-dimensional QED Lagrangian features interactions 
quartic in the field strength, which might contribute gauge dependent terms. So additional 
tree-level identities might be required.

A further extension of our results might include non-linear gauges. A discussion of the 
cancellation identities of QED in the 't Hooft-Veltman gauge \cite{'tHooft:1973pz} will follow 
elsewhere.

From the combinatorial proof of the Ward identities, 
\cite{'tHooft:1971fh,'tHooft:1972ue,Lautrup:1977ty,Cvitanovic:1977qu,Kreimer:2012jw,Sars:2015phd} 
it is reasonable to expect similar cancellations for the quark propagator in QCD. However, one can 
not expect a linear Dyson-Schwinger equation to describe the gauge dependence, rather, we 
anticipate that the scalar propagator of the skeleton graph obtains ghost self-energy insertions and 
that the ghost-gluon vertex induces additional skeletons, which potentially relate to the 
appearance of higher-order Casimir operators.

Another exciting project will be to investigate cancellation identities in the context of massive 
gauge bosons and to examine the role of the Goldstone and Higgs bosons in models with spontaneously 
broken symmetry. However, these questions are clearly beyond the scope of this letter and deserve 
further investigation.

\section*{Acknowledgments}
This letter would not have been possible without several stimulating discussions with 
D.J.~Broadhurst and J.A.~Gracey. Special thanks goes to A.~Vogt for useful discussions and 
pointing out the results of the papers \cite{Chetyrkin:1999pq,Ruijl:2016pkm}. H.K.\ thanks the 
University of Liverpool for hospitality as well as E.~Panzer for his hospitality and encouragement 
to finish this article. Further thanks to K.~Jones for helpful comments on the manuscript. The 
figures in this letter were created with {\sc jaxodraw} \cite{Binosi200476} and the {\sc axodraw} 
packages \cite{Vermaseren:1994je,Collins:2016aya}. This work was partially supported by a DAAD 
scholarship.

\bibliographystyle{elsarticle-num}
\bibliography{cancel}

\end{document}